\documentclass[prc,aps,superscriptaddress,showpacs,nofootinbib,twocolumn]{revtex4}
\usepackage{amssymb}
\usepackage[dvips]{graphicx}
\usepackage[english]{babel}
\usepackage{indentfirst}
\usepackage{amsxtra}
\usepackage{amsmath}
\usepackage{supertabular}
\usepackage{multirow}
\usepackage [mathcal]{eucal}
\begin{document}
\title {\bf Two-neutron transfer probabilities and spatial-localization effects at the drip line}

\author{Marcella Grasso}
\affiliation{Institut de Physique Nucl\'eaire, IN2P3-CNRS, Universit\'e Paris-Sud, 
F-91406 Orsay Cedex, France}

\begin{abstract} 
We examine Cr isotopes at the drip line, where surface effects related to the existence of a weakly bound $s1/2$ state are known to be important and tightly connected with the pairing phenomenon (anti-halo effect). For these weakly-bound isotopes, we evaluate the ground state to ground state two-neutron transfer probabilities within a mean-field-based approach. 
An important part of the discussion is devoted to the analysis of several procedures that can be employed to constrain the parameters of a phenomenological pairing interaction. The parameters are first adjusted to reproduce the experimental gaps evaluated with the five-point formula. This choice has 
however some consequences on the evolution of the pairing correlations along the isotopic chains and, in particular, at shell closures. Other procedures are then followed (adjustment on a theoretical pairing gap at mid-shell and on the two-neutron separation energies). 

For the transfer probabilities, 
 we discuss the effects associated to different  choices of the spatial localization of the pairing interaction.
We indicate that the analysis of 
pair-transfer reactions for such cases (where the last bound state is a low-$l$ state in a weakly bound nucleus) may improve our understanding of two aspects: the spatial distribution of pairing correlations in nuclei and the general problem of the persistence of pairing at the drip lines. 

\end{abstract} 

\vskip 0.5cm  \maketitle 
% 

%-----------------------------------------------------------------------
\section{Introduction}

Pairing correlations and pair-transfer reactions are intimately related  as indicated by several theoretical studies \cite{ripka,broglia,oertzen}.
Multi-nucleon transfer experiments are currently performed in several facilities around the world 
\cite{Cor11,ganil,canberra}. The forthcoming extensions of this type of measurements to more exotic and weakly bound isotopes will open new opportunities for a deeper understanding of the pairing phenomenon in nuclei where surface effects are important (diffuseness of the surface, neutron skins, etc.). These future experiments may indeed provide a better insight into the spatial localization of the pairing correlations and of the Cooper pairs in nuclei.  
Recently, it has been shown by several authors that the analysis of the addition or removal modes  associated to the transfer from a nucleus $A$ to a nucleus $A\pm2$ may give indications on the nature of the pairing interaction in terms of its spatial localization \cite{Mat11,matsuo,khan,pllumbi,grasso12}. In particular,  ($p,t$) transfer reactions for very neutron-rich Sn isotopes have been suggested in Ref. \cite{pllumbi} as possible candidates to investigate the volume and/or surface spatial properties of the pairing interaction. This indication remains however qualitative: in spite of the fact that the structure calculation for the evaluation of the form factors were fully microscopic and self-consistent in Ref. \cite{pllumbi}, the zero-range distorted-wave Born approximation employed for the reaction calculations did not allow the authors to draw quantitative conclusions on the obtained cross sections. 

In Ref. \cite{grasso12} this issue has been discussed again by focusing only on transitions from the ground state of the nucleus $A$ to the ground state of the nucleus $A\pm2$. Several formula to calculate the associated addition and removal transition probabilities have been considered and the corresponding results compared. Enhancement effects in the transition probabilities  observed for some Sn isotopes (related to the use of a surface-peaked pairing interaction and also found in Ref. \cite{matsuo}) have been analyzed. In Ref. \cite{matsuo} these effects are described in terms of radial integrals (used to evaluate the transition probabilities) where low-$l$ wave functions contribute for some very neutron-rich isotopes. These effects are amplified when surface-peaked interactions are employed because  low-$l$ wave functions have a strong localization at the surface. In Ref. \cite{grasso12} the same enhancement effects in the transition probabilities have been interpreted in a complementary way within the canonical-basis picture. In this scheme, the transition probabilities are not calculated as radial integrals, but as sums of occupation probabilites. The enhancement effects related to surface-peaked interactions are due in this representation to a more pronounced fragmentation of the canonical states around the Fermi energy. 

A crucial aspect to consider in this type of analyses is related to the criteria that are chosen to constrain the pairing interaction. In Refs. \cite{khan,pllumbi,grasso12} the parameters of the different pairing interactions are adjusted to reproduce the two-neutron separation energies for those even Sn isotopes in the nuclear chart where masses are known experimentally. In other calculations (for instance, in Ref. \cite{margue}) the parameters have been fitted to reproduce the experimental gaps (evaluated with the three-point or the five-point formula \cite{duguet}) with a theoretical gap calculated as 
\begin{equation}
\Delta_q = \frac{\int dr \rho_q(r) \Delta_q(r)}{\int dr \rho_q(r)},
\end{equation}
or
\begin{equation}
\tilde{\Delta}_q = \frac{\int dr \tilde{\rho}_q(r) \Delta_q(r)}{\int dr \tilde{\rho}_q(r)},
\end{equation}
where $\Delta_q(r)$ is the pairing potential for neutrons ($q=\nu$) or protons ($q=\pi$); $\rho_q(r)$ and $\tilde{\rho}_q(r)$ are the nucleonic particle and anomalous densities, respectively, defined within the Hartree-Fock-Bogoliubov (HFB) framework. In both the previous procedures, that of Refs.   
\cite{khan,pllumbi,grasso12} and that of Ref. \cite{margue}, 
the adjustment of the parameters is done only on even nuclei. 
As far as the second procedure is concerned, two aspects should be emphasized. First, there is an ambiguity related to the choice of the two possible quantities, Eq. (1) or Eq. (2), to evaluate the gap in the theoretical calculations. The two quantities may be quite different 
one from the other for some nuclei and for some specific choices of the pairing interaction as shown, for example, in Ref. \cite{grasso12}. Second,  
$\Delta_q $ and $\tilde{\Delta}_q $
 are both calculated in a different way with respect to the 
experimental values obtained with the three- or the five-point formula. The comparison 
between such experimental and theoretical values is thus questionable.

A third example is given in    
Ref. \cite{scamps} where the parameters are adjusted for stable Ca isotopes to reproduce the experimental gaps by using the five-point formula \cite{duguet}. The five-point formula is 
also used in the theoretical 
calculations and odd nuclei are included by using the blocking approximation. 
The comparison between the theorerical and the experimental values is now meaningful because the same quantities are evaluated in both cases. 
By using this prescription, negligible discrepancies are found in the transfer transition probabilities for Ca isotopes for differently localized pairing interactions (surface-peaked, surface/volume mixed and volume). However, the calculations are done only for stable nuclei and the exotic regions  are not explored. Furthermore, it could  be interesting to check what this type of fitting procedure implies 
in the theoretical predictions for the 
 pairing correlations along the isotopic chains and, in particular, at shell closures.

In this article we address two main issues: i) analyzing how much the predictions (and the surface effects) related to pair-transfer reactions are affected by the procedure followed to adjust the parameters of the pairing interaction. For a surface-peaked, a mixed and a pure volume interaction, we adjust first the parameters to reproduce the experimental pairing gaps calculated with the five-point formula. Odd nuclei are included in the adjustment procedure and they are treated with the blocking approximation in the HFB calculations. 
To fit the parameters of the neutron pairing interaction, we choose Ca isotopes, from $^{22}$Ca to $^{26}$Ca. 
We analyze in detail what this procedure implies for the anomalous densities and calculate the corresponding trends of the two-neutron separation energies to check whether they are in a reasonable agreement with the experimental values (this was the fitting criterium used in Refs. \cite{khan,pllumbi,grasso12}). 
We then generate two other sets of parameters: one obtained to have the same value of $\Delta_{\nu}$, Eq. (1), for the mid-shell nucleus $^{44}$Ca; the other obtained to reproduce the two-neutron separation energies from $^{42}$Ca to $^{48}$Ca.
ii) By using the different procedures to adjust the pairing interaction, we consider a case where surface effects are expected to be particularly amplified. In recent studies
 \cite{hagino1,hagino2} large odd-even staggering effects have been found in reaction cross sections of unstable Ne and Cr nuclei close to the neutron drip line. Such strong staggering effect has been related to the existence of low-$l$ weakly bound states in drip line nuclei and to the associated anti-halo effect generated by the pairing interaction \cite{bennaceur,grasso,rotival}.  
In Ref. \cite{hagino2} the previous predictions found by Hamamoto et al. \cite{hamamoto1,hamamoto2}
(the pairing gap is expected to disappear for weakly bound states) are shown to be an artificial result related to the employed approximations. This result was actually found in Refs. \cite{hamamoto1,hamamoto2} because the calculations were not self-consistent in the determination of the Fermi energy $\lambda$. This issue is associated to the more general problem of the persistence of the pairing gap in weakly bound superfluid systems, such as nuclei at the drip lines as well as nuclear clusters in the crust of neutron stars \cite{schuck,pastore}. 

The special cases discussed in Refs. \cite{hagino1,hagino2} may be important to unambigously disentangle among the different possible spatial localizations of the pairing interaction. In this work we analyze one of these cases, in the Cr isotopic chain, and we compute the ground state to ground state pair-transfer transition probabilities at the neutron drip line to check the sensitivity of the results on the choice of the interaction. The problem of the persistence of pairing correlations at the drip lines is shown to be connected with the transfer-probability trends.      

The article is organized as follows. 
In Sec. II the choice of the procedure employed to adjust the parameters of the pairing interaction is discussed. Different spatially localized neutron pairing interactions are generated for stable open-shell Ca isotopes. 
In Sec. III these different pairing interactions are used to calculate the ground state to ground state two-neutron transfer probabilities in Cr isotopes at the drip line, where the $3s1/2$ neutron orbital is the last (and weakly bound) state. 
In Sec. IV conclusions are reported.   

\section{Adjustment of the pairing interactions}

We perform HFB calculations with the Skyrme parametrization SkM* \cite{skm} in the mean-field channel. Odd nuclei are treated with the blocking approximation. 
A zero-range density-dependent pairing interaction $V_{pair}$ is employed in the pairing channel, with 
\begin{equation}\label{eq:vpair}
 V_{pair}({\bf r_1},{\bf r_2})=V_0\left[1-\eta\left(\frac{\rho({\bf R})}{\rho_0}\right)^\alpha\right]
 \delta\left({\bf r_1}-{\bf r_2}\right),
\end{equation}
where ${\bf R}=({\bf r_1}+{\bf r_2})/2$, $\alpha=1$ and $\rho_0=$ 0.16 fm$^{-3}$. With a cutoff equal to 60 MeV in quasiparticle energies and a maximum value of $j$ equal to 15/2, the parameter $V_0$ was adjusted in Refs. \cite{khan,pllumbi,grasso12} to reproduce the known  two-neutron separation energies of Sn isotopes . By using this prescription, it was checked that the proton pairing correlations were equal to zero for all the considered isotopes and that the neutron pairing correlations were equal to zero for the doubly magic nuclei along the isotopic chain, as should be expected for calculations done within the mean-field-based HFB framework. 

First, we change the prescription and adjust the intensity of the pairing interaction $V_0$ 
on the experimental neutron pairing gaps $\Delta^{(5)}$ evaluated with the five-point formula: 

\begin{widetext}
\begin{equation}
\Delta^{(5)} (Z,N) = - \frac{(-1)^N}{8} [E(Z,N+2)-4E(Z,N+1) 
+ 6E(Z,N)-4E(Z,N-1)+E(Z,N-2)], 
\end{equation}
\end{widetext}
where $E(Z,N)$ is the binding energy of the nucleus. We choose the five-point formula instead of the three-point one to have more regular trends to be fitted and to avoid the odd-even staggering effect. 
In Sec. III we will analyze Cr isotopes and calculate two-neutron transfer probabilities at their  (theoretically predicted) neutron drip line. 
To perform these calculations, 
we need a suitable neutron pairing interaction. 
In principle, we should do the adjustment of the parameters for the Cr isotopic chain by using the known experimental masses, that is, from $^{46}$Cr to 
$^{62}$Cr. However, the experimental $\Delta^{(5)}$ values have a very irregular trend for such Cr isotopes. 
These irregularities may be related to the fact that some isotopes are deformed and also to the presence of a proton pairing for these nuclei.  

%\begin{figure}[htb]
%\begin{center}
%\includegraphics[width=8cm]{gapCr.eps}
%\end{center}
%\caption{Experimental gaps for odd and even Cr isotopes, calculated with the five-point formula. 
%}
%\end{figure}

\begin{figure}[htb]
\begin{center}
\includegraphics[width=8cm]{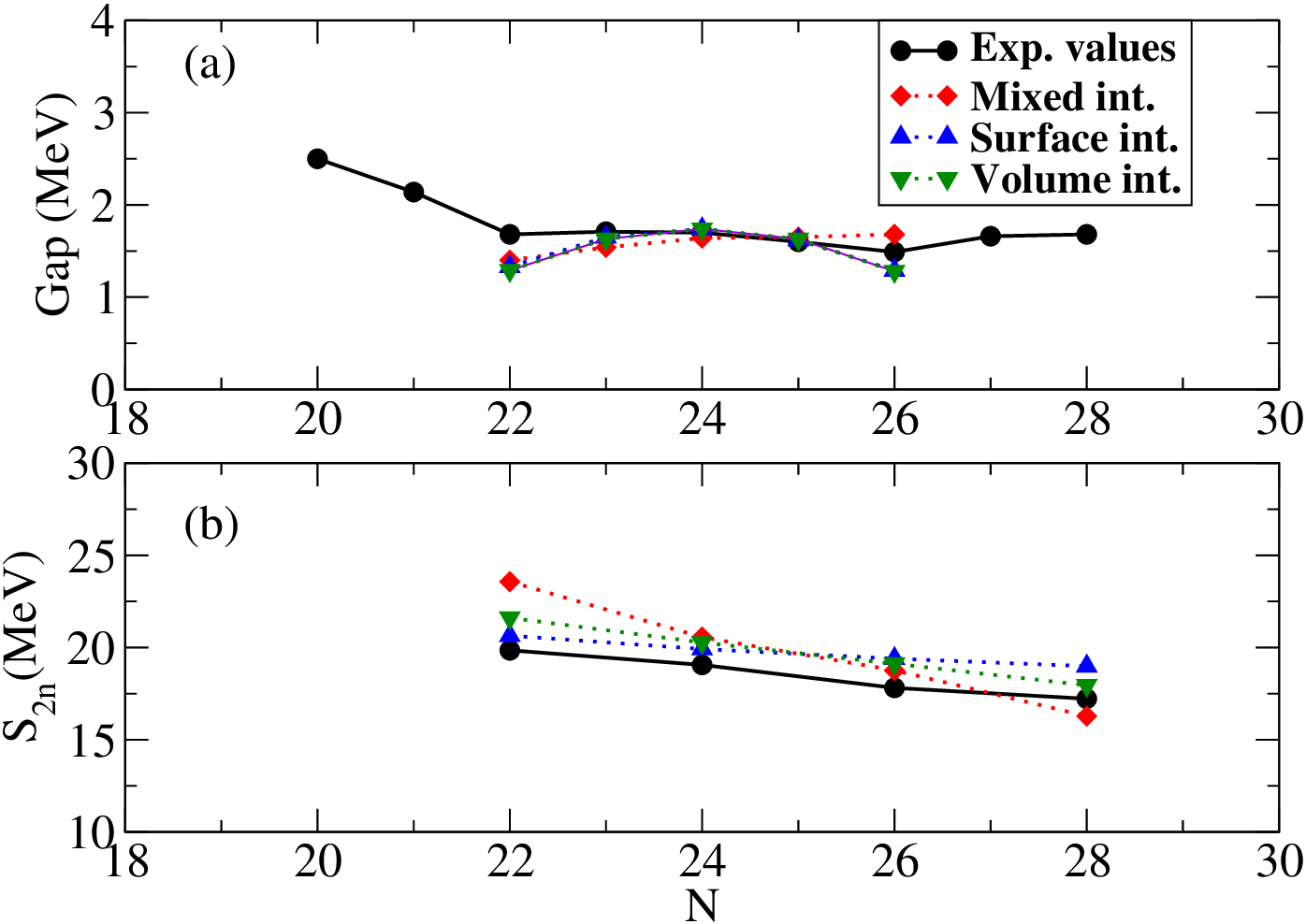}
\end{center}
\caption{ (Color online) 
(a) Neutron gaps $\Delta^{(5)}$ obtained with the five-point formula for Ca isotopes. (b) Two-neutron separation energies for Ca isotopes. The calculations are performed with the parameters of Table I. 
}
\end{figure}

\begin{figure}[htb]
\begin{center}
\includegraphics[width=8cm]{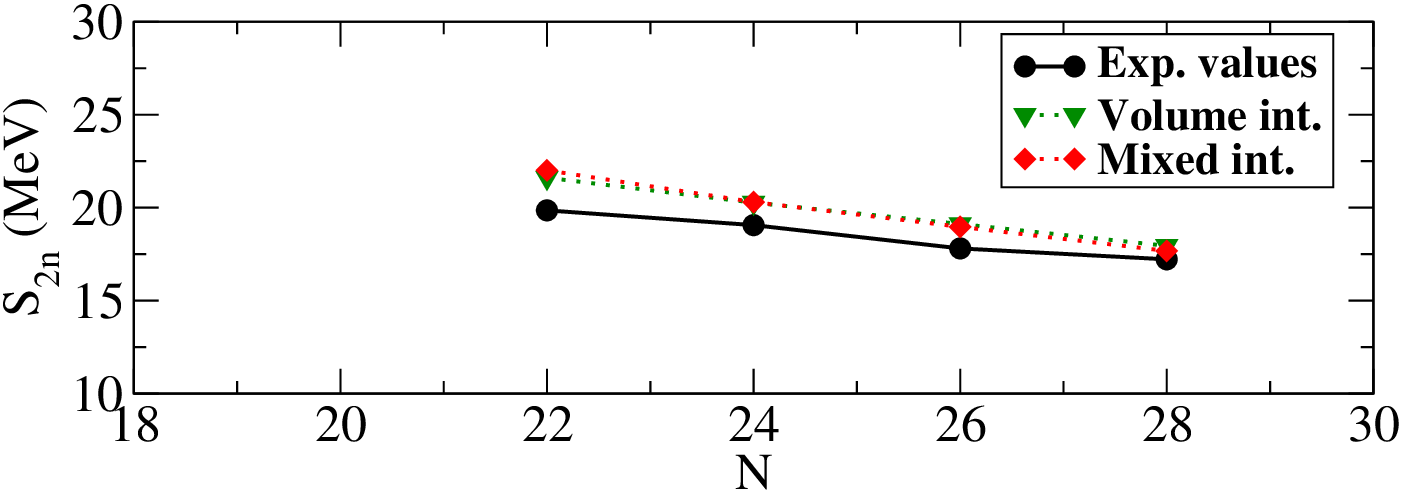}
\end{center}
\caption{ (Color online) 
Two-neutron separation energies for Ca isotopes corresponding to the interactions listed in Table II. 
}
\end{figure}

To simplify the fitting procedure for the neutron pairing, we take a more regular case by choosing isotopes with a similar number of $Z$ and where the proton pairing is negligible. We choose Ca isotopes where the trend of $\Delta^{(5)}$ is indeed quite regular, as can be seen in Fig. 1.  
We adjust the different pairing interactions to reproduce  the trend of 
$\Delta^{(5)}$
for both odd and even nuclei, from 
$^{42}$Ca to $^{46}$Ca. We keep out from the fit the nuclei close to the magic isotopes $^{40}$Ca and $^{48}$Ca. It is known that for these nuclei the value of the gap 
$\Delta^{(5)}$
increases owing to spurious effects that are not related to the pairing correlations but to the shell structure. 
Our objective is to reproduce at best the trend of the gap $\Delta^{(5)}$, with a special care 
for the mid-shell nucleus $^{44}$Ca.
To be sure that the proton pairing is negligible for all the isotopes under study, we settle the intensity $V_0$ of the proton pairing interaction at the value -150 MeV fm$^{-1}$ for all the examined cases.  
We plot in Fig. 1 (a) the corresponding values of $\Delta^{(5)}$ and, in panel (b), the trends of the two-neutron separation energies $S_{2n} (Z,N)=E(Z,N) - E(Z, N-2) $, for three interactions, surface-peaked ($\eta=1$), mixed ($\eta=0.5$) and volume ($\eta=0$). 
For the separation energies, we consider the isotopes from $^{42}$Ca to $^{48}$Ca. We observe that 
we have succeeded in having almost the same value of 
$\Delta^{(5)}$ for the nucleus $^{44}$Ca, with the three 
pairing interactions. This value is in good agreement with the experimental data. Globally, the two-neutron separation energies are not in very good agreement with the experimental trends, 
especially for the mixed interaction, where the discrepancies are more pronounced and even the slope 
is strongly different.  

\begin {table} 
  
\begin{tabular}{cccc}
\hline
$\eta$ & $V_0$ (MeV fm$^{-1}$) & $\Delta_{\nu}$ (MeV) &  $\Delta^{(5)}$ (MeV)\\
& & $^{44}$Ca & $^{44}$Ca \\
\hline 
\hline
0 (Volume) & -185 & 1.52 & 1.74 
\\  
0.5 (Mixed) & -376 & 3.34 & 1.64 \\
1 (Surface) & -295 & 0.55 & 1.74 \\
\hline
\end{tabular}
\caption {Second column: values of the strength $V_0$ of the neutron pairing interaction adjusted to reproduce the experimental trends of $\Delta^{(5)}$ calculated with the five-point formula, for three different pairing interactions, volume, mixed and surface--peaked (different values of $\eta$). In the third and fourth columns the corresponding values of the neutron $\Delta_{\nu}$ and $\Delta^{(5)}$, calculated with Eq. (1) and with the five--point formula, respectively, are reported for the nucleus $^{44}$Ca.}
\end{table}

In Table I, we list the corresponding values of $V_0$ for the surface-peaked ($\eta=1$), the mixed ($\eta=0.5$) and the volume ($\eta=0$) 
pairing interactions and the corresponding values of the neutron $\Delta_{\nu}$, evaluated 
by using Eq. (1), for the nucleus $^{44}$Ca. This is done to check whether the adopted criterium to reproduce the experimental value of   $\Delta^{(5)}$ 
also leads to similar values of $\Delta_{\nu}$. We clearly see that this is not true:
the values of $\Delta_{\nu}$ for $^{44}$Ca are extremely different for the three pairing interactions. 
We have checked then that, if we want to have, for example, a value of  of $\Delta_{\nu}$ of $\sim 1.5$ MeV 
for $^{44}$Ca,  
we may keep the same volume interaction as in Table I, but we have to change the values of $V_0$ for the other pairing interactions by reducing the strength for the mixed case and by increasing the strength for the surface-peaked case. The corresponding values are reported in Table II. 
\begin{table} 
\begin{tabular}{cccc}
\hline
$\eta$ & $V_0$ (MeV fm$^{-1}$) & $\Delta_{\nu}$ (MeV)& $\Delta^{(5)}$ (MeV) \\
& & $^{44}$Ca &   $^{44}$Ca \\
\hline 
\hline
0  (Volume) & -185 & 1.52 & 1.74 
\\  
0.5 (Mixed) & -280 & 1.56 & 0.38 \\
1  (Surface) & -510 & 1.56 &   \\
\hline
\end{tabular}
\caption{Second column: values of the strength $V_0$ of the neutron pairing interaction adjusted to have in all cases $\Delta_{\nu}$ $\sim$ 1.5 MeV for the nucleus $^{44}$Ca. In the third and fourth columns the corresponding values of the gap calculated with Eq. (1) and with the five-point formula are reported for the nucleus $^{44}$Ca.}
\end{table}
If we use now these strengths, we obtain for the nucleus 
$^{44}$Ca values of $\Delta^{(5)}$ equal to 1.74 and 0.38 MeV for the volume and mixed interactions, respectively (third column of Table II). The calculations for the odd and even isotopes close to the nucleus $^{44}$Ca do not converge numerically for the surface-peaked interaction of Table II and the corresponding values of $\Delta^{(5)}$ and of $S_{2n}$ cannot thus be calculated. By comparing the volume and mixed cases in Table II, we  
conclude that, if we adopt the criterium of having similar values of $\Delta_{\nu}$ at mid-shell, then the strength of the mixed interaction has to be strongly reduced and this provides a too low value of 
$\Delta^{(5)}$, that is far from the experimental value. 

The separation energies corresponding to the interactions of Table II are plotted in Fig. 2 (with the exception of the surface-peaked interaction). 
We observe that the slope of the $S_{2n}$ values corresponding to the mixed interaction has been corrected with respect to that of Fig. 1: the agreement with the experimental trend has been globally improved.   

The anomalous densities corresponding to all the interactions of Tables I and II are plotted for the nucleus $^{44}$Ca in Fig. 3. The three upper panels refer to the interactions of Table I, whereas the three lower panels refer to the interactions of Table II. We observe that the criterium of constraining the pairing by using the  experimental values $\Delta^{(5)}$ leads to very different profiles for the anomalous densities for the nucleus $^{44}$Ca for the three different choices of the pairing interaction. This does not occur when the prescription of having a similar value of $\Delta_{\nu}$ for $^{44}$Ca is adopted. 

\begin{figure}[htb]
\begin{center}
\includegraphics[width=8cm]{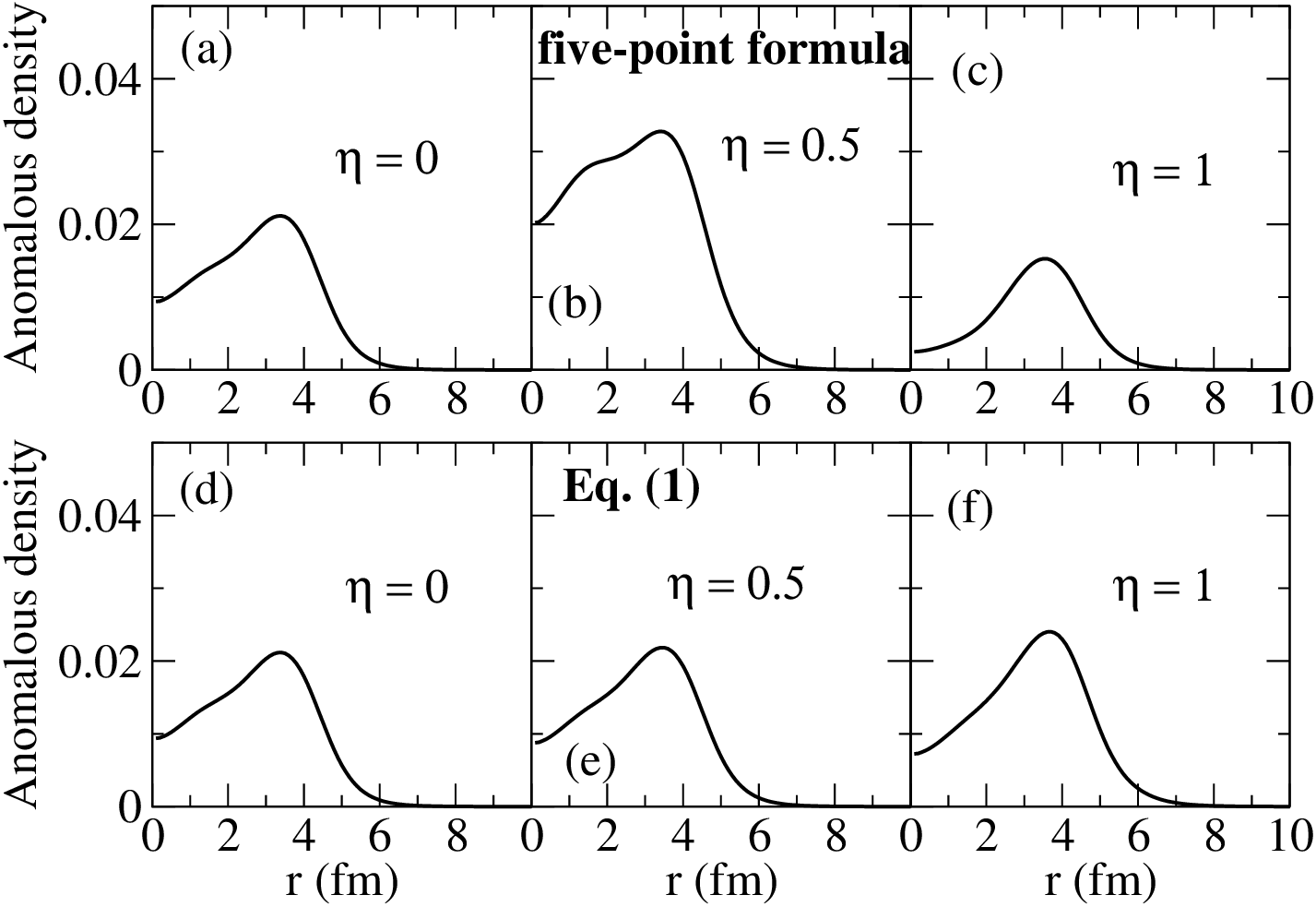}
\end{center}
\caption{
Anomalous densities (in units of fm$^{-3}$) for the nucleus $^{44}$Ca. In panels (a), (b) and (c) the densities correspond to the interactions of Table I. In panels (d), (e) and (f) the densities correspond to the interactions of Table II. 
}
\end{figure}

\begin{figure}[htb]
\begin{center}
\includegraphics[width=8cm]{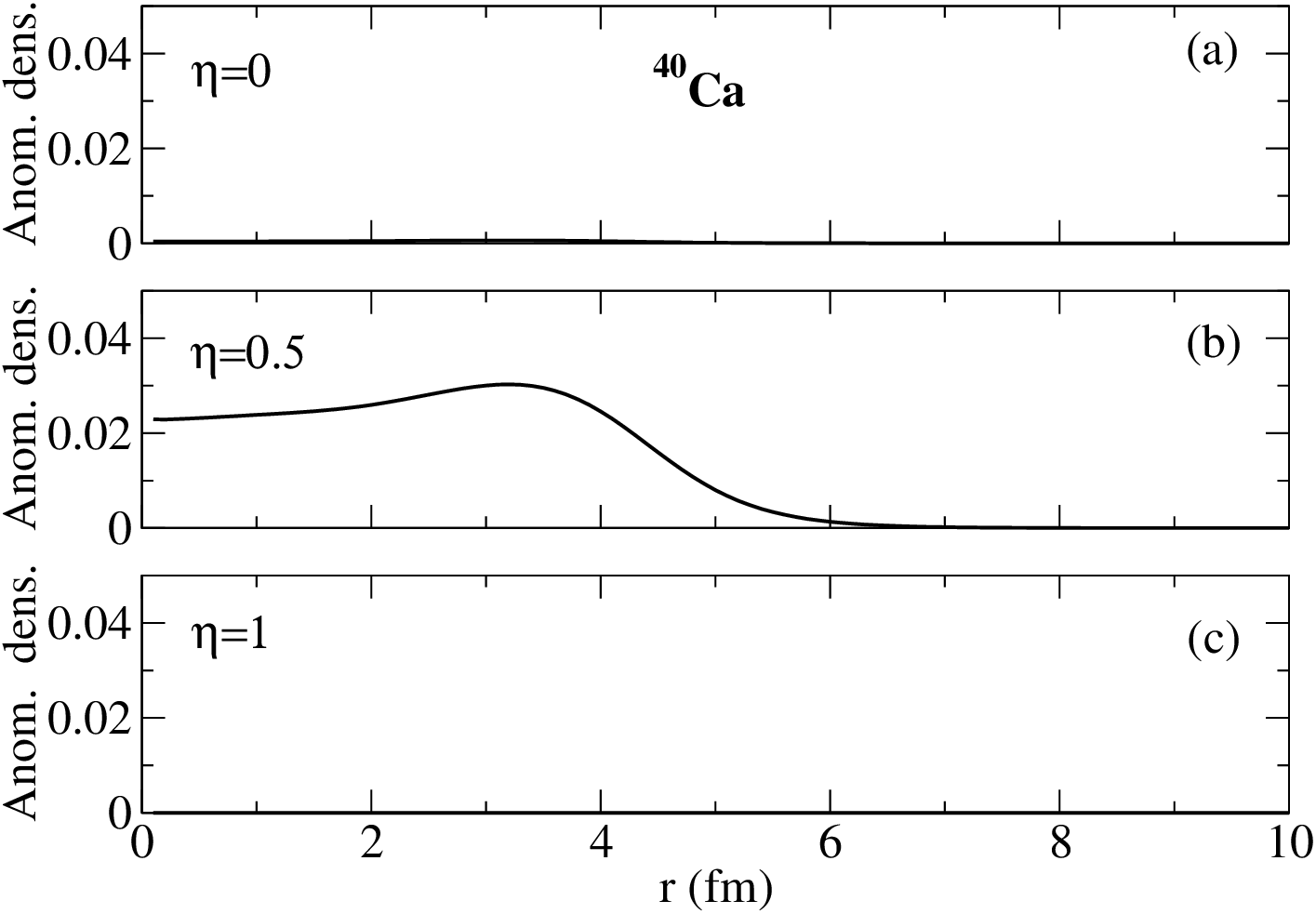}
\end{center}
\caption{Anomalous densities (in units of fm$^{-3}$) for the nucleus $^{40}$Ca. The densities correspond to the interactions of Table I. 
}
\end{figure}

There is an additional aspect that should be addressed. When the prescription of reproducing the experimental values of   $\Delta^{(5)}$ for open-shell nuclei is used, it is not garanteed that the pairing correlations are quenched at shell closures. In Fig. 4 we plot the neutron anomalous densities for the magic nucleus 
$^{40}$Ca, where pairing correlations are expected to disappear in HFB calculations. The profiles are shown for the three pairing interactions of Table I. 
We observe that the anomalous density is negligible for the volume interaction (a) and  for the surface-peaked interaction (c). For the mixed interaction (b), however, the anomalous density is very large. We have checked that in this case the value of $\Delta_{\nu}$ is 
$\sim$ 3 MeV. The pairing interaction is too intense to provide a closed-shell $^{40}$Ca. On the other hand, if we reduce the strength of the pairing interaction to predict a magic nucleus at $N=20$, the 
value of $\Delta^{(5)}$ becomes too weak for the open-shell nuclei (see Table II).
This particular behavior found for the mixed interaction can be seen also in Figs. 5 and 6, where the occupation numbers of the neutron $s1/2$, $d3/2$ and $f7/2$ quasiparticle states are reported for the nucleus $^{40}$Ca, up to a quasiparticle energy of 10 MeV. Fig. 5 refers to the mixed interaction and Fig. 6 to the surface-peaked interaction of Table I. We see that the occupation numbers are equal to 1 for the occupied states only for $\eta=1$. For the mixed interaction, the occupied states have an occupation number smaller than 1 and the $f7/2$ orbital starts being occupied. The ground state of $^{40}$Ca is thus far from being uncorrelated as should be expected in HFB calculations. 

Several aspects shown in this work seem to indicate that the fitting criterium employed to provide the parameters of Table I may lead to questionable results: i) the anomalous densities obtained with different (in the spatial localization) pairing interactions  may be very different for a superfluid nucleus (in our case $^{44}$Ca) [panels (a), (b) and (c) of Fig. 3]; ii) pairing correlations may be active even for those nuclei that are expected to be closed shell (in our case $^{40}$Ca), especially for some choices of the pairing interaction (here, for mixed pairing interactions); iii) last but not least, we observe that the trend of the parameters is not the same as that of Tables II and III. What is physically expected is that the strength of the interaction should increase when going from a volume to a surface--peaked interaction. This behavior is not observed in Table I.  

Let us now change the procedure and adjust the parameters to reproduce at best the trend of the separation energies. The obtained parameters are reported in Table III and the separation energies are shown in Fig. 7. We have checked that the nucleus $^{40}$Ca is now predicted almost closed-shell in all cases. The values of $\Delta_{\nu}$
 and $\Delta^{(5)}$ are also included in the table for the nucleus $^{44}$Ca. 
The numerical calculations for the surface-peaked interaction do not converge for the odd nuclei close to 
$^{44}$Ca in this case and the corresponding value of $\Delta^{(5)}$ cannot be computed. 
 The three values of $\Delta_{\nu}$ are very similar. 
The two values of $\Delta^{(5)}$  for $\eta=0$ and 0.5 are very different.

\begin{table} 
\begin{tabular}{cccc}
\hline
$\eta$ & $V_0$ (MeV fm$^{-1}$) & $\Delta_{\nu}$ (MeV) & $\Delta^{(5)}$ (MeV) \\
& & $^{44}$Ca & $^{44}$Ca \\
\hline 
\hline
0 (Volume) & -170 & 1.26 & 1.49  
\\  
0.5 (Mixed) & -250 & 1.23 & 0.48 \\
1 (Surface) & -460 & 1.24 &   \\
\hline
\end{tabular}
\caption{Second column: values of the strength $V_0$ of the neutron pairing interaction adjusted to reproduce at best the trend of the separation energies. In the third and fourth columns, the corresponding values of the gap calculated with Eq. (1) (third column) and with the five-point formula (fourth column) are reported for the nucleus $^{44}$Ca.}
\end{table}
 
There are then two possible attitudes that can be adopted to adjust the parameters of the pairing interaction: (i) reproducing the experimental trends of the two-neutron separation energies $S_{2n}$. This allows us, at the same time, to predict reasonable closure shells at magic nuclei. With this procedure, the resulting values of $\Delta_{q}$ practically do not differ. The resulting values of $\Delta^{(5)}$ are strongly different. (ii) Reproducing the $\Delta^{(5)}$ values, being aware, however, that the shell closures and the two-neutron separation energies may be poorly reproduced, owing to too strong pairing intensities in some cases. 

In Ref. \cite{scamps} the procedure (ii) has been employed and negligible differences have been found in transfer probabilities for stable nuclei and for three different localizations of the pairing interaction. In next Section we use the same procedure by exploring more exotic cases, at the drip line of Cr isotopes. 

\begin{figure}[htb]
\begin{center}
\includegraphics[width=8cm]{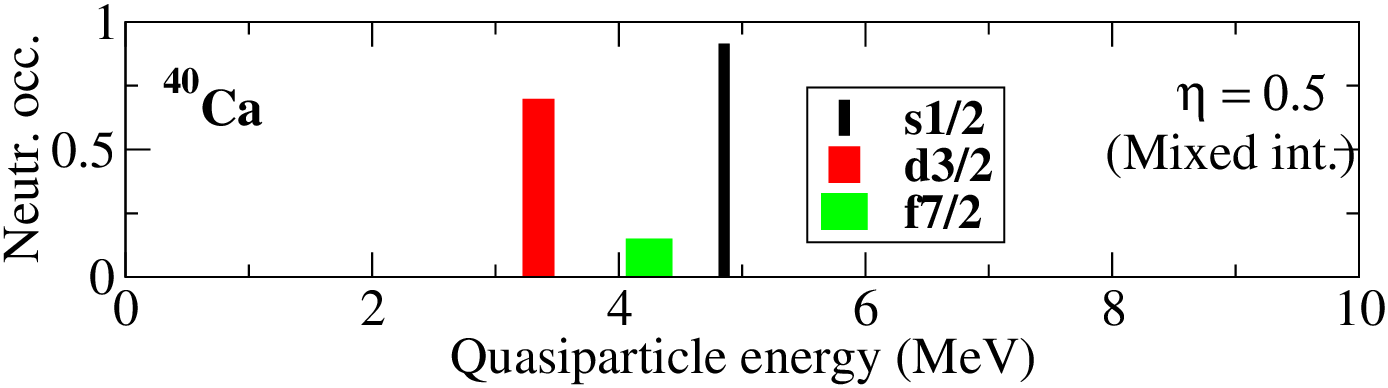}
\end{center}
\caption{(Color online) Occupation numbers of the neutron states $s1/2$, $d3/2$ and $f7/2$ as a function of the quasiparticle energy, up to 10 MeV, for the nucleus $^{40}$Ca. The two states with a larger occupation probability correspond to the orbitals $2s1/2$ and $1d3/2$. The orbital $1f7/2$ has an occupation different from zero. The mixed interaction of Table I is used.  
}
\end{figure}

\begin{figure}[htb]
\begin{center}
\includegraphics[width=8cm]{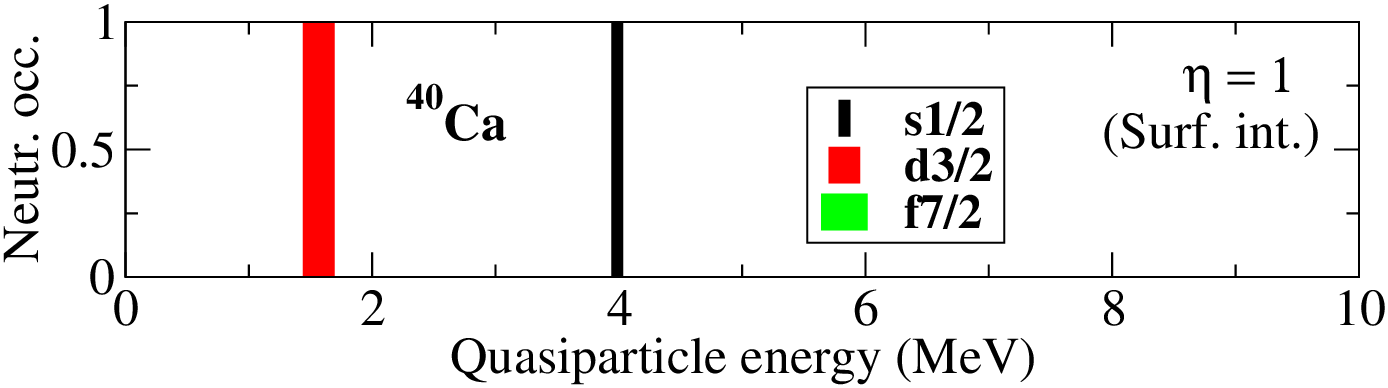}
\end{center}
\caption{(Color online) Same as in Fig. 6 but for the surface-peaked interaction of Table I. In this case, the 
$2s1/2$ and $1d3/2$ orbitals have occupation numbers equal to 1 and the. The $1f7/2$ orbital is empty. }
\end{figure}

\begin{figure}[htb]
\begin{center}
\includegraphics[width=8cm]{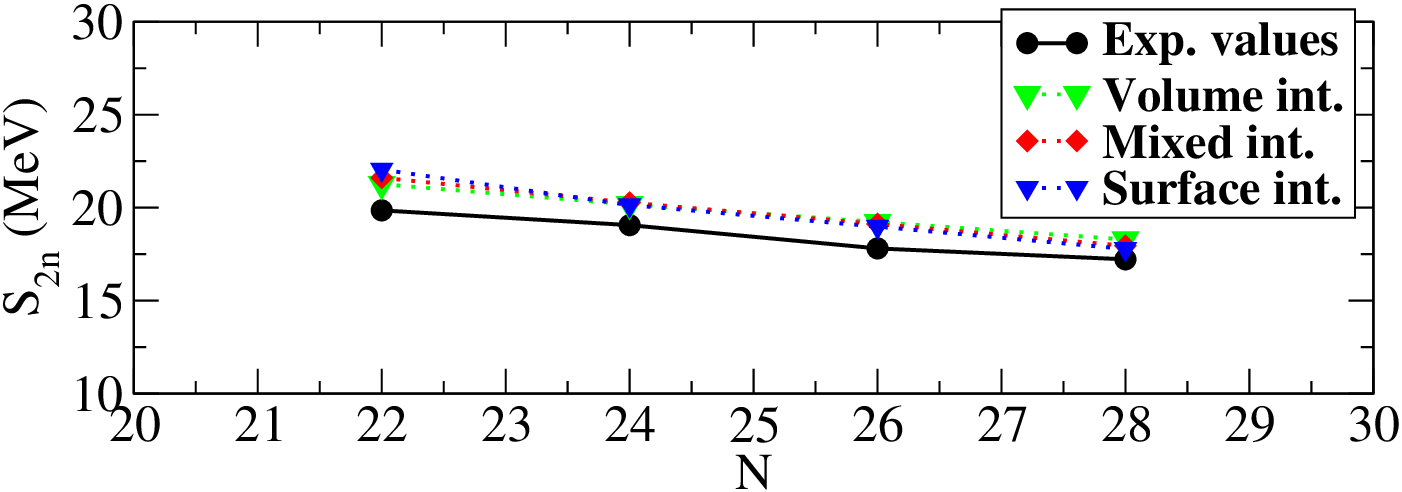}
\end{center}
\caption{ (Color online) 
Two-neutron separation energies for Ca isotopes, from $^{42}$Ca to $^{48}$Ca, obtained with the parameters of Table III.}
\end{figure}

\section{Pair-transfer probabilities for Cr isotopes at the drip line}

We employ the parameters adjusted in the previous Section. The same parameters are kept also for the pairing interaction (even if the present isotopes are not closed-shell in protons) to better isolate the effets coming only from the neutron pairing. 
In Ref. \cite{hagino2} the Cr drip-line nucleus is $^{76}$Cr ($N=52$) where the last bound state is the 
$3s1/2$ neutron orbital. Calculations are done there with a Woods-Saxon potential in the mean-field channel. We use here the Skyrme parametrization SkM*. The Hartree-Fock drip-line Cr isotope is in our case the 
nucleus $^{82}$Cr ($N=58$). The last bound occupied state is also the  $3s1/2$ neutron orbital, but 
the orbital $2d3/2$ is found here between the states $1g9/2$ and $3s1/2$. This explains why the drip line is shifted by six isotopes. 

The removal and addition ground state (GS) $\rightarrow$ GS pair-transfer amplitudes are written as,  
\begin{equation}
T^{\rm Rem} = \langle {\rm GS}_{A-2} | \Psi_{q'} ({\bf{r_1}},-\sigma_1) \Psi_q ({\bf{r_2}},\sigma_2) | {\rm GS}_{A} \rangle, 
\label{rem}
\end{equation}
and
\begin{equation}
T^{\rm  Add} = \langle {\rm GS}_{A+2} | \Psi^{\dag}_{q'} ({\bf{r_2}},\sigma_2) \Psi^{\dag}_q ({\bf{r_1}},-\sigma_1) | {\rm GS}_{A} \rangle, 
\label{add}
\end{equation}
respectively, 
where $\sigma$ and $q$ represent spin and isospin, respectively. The matrices are assumed diagonal in isospin, that is $q'=q$. 
The states $| {\rm GS}_{A} \rangle$ and $| {\rm GS}_{A \pm 2} \rangle$ correspond to the ground states 
of the initial and of the final nucleus
 of the transfer reaction. 

In the approximation of Ref. \cite{matsuo} the pair-transfer probabilities are calculated with the amplitudes
\begin{equation}
T^{Rem}=T^{Add}=-\frac{1}{4\pi r^2} 
\sum_{nlj}u^A_{nlj}v^A_{nlj}=\tilde{\rho}(r). 
\end{equation}
They are thus written as: 
\begin{equation}
P^{Rem}=P^{Add}=|\int dr \sum_{nlj}u^A_{nlj}v^A_{nlj}|^2.  
\end{equation}
The improved formula introduced in Ref. \cite{grasso12} leads to: 
\begin{equation}
P^{Rem}(A+2)=P^{Add}(A), 
\end{equation}
and
\begin{equation}
P^{Add}(A)=|\int dr \sum_{nlj}u^A_{nlj}v^{A+2}_{nlj}|^2.  
\end{equation}
We calculate the addition probabilities associated to the transfers $^A$Cr $\rightarrow$ $^{A+2}$Cr, starting from the nucleus $^{74}$Cr where the orbital $1g9/2$ is fully occupied up to the nucleus 
$^{82}$Cr, where the neutron state $3s1/2$ contributes. We have checked that, for all the employed interactions, $|E(Z,N)| > |E(Z,N-2)|$, to be sure that the HFB two-neutron drip line is not predicted between $^{74}$Cr and $^{82}$Cr. The $3s1/2$ neutron orbital 
is contributing only for the nucleus $^{82}$Cr. 

In Figs. 8 and 9 the transfer probabilities are plotted for the volume (a), mixed (b), and surface-peaked (c) interaction, for the parameters of Tables I (Fig. 8) and II (Fig. 9). The two equations (8) and (10) (improved form) are used. Let us consider first Fig. 8 (parameters of Table I). The strengths are strongly different, according to the used pairing interaction. In particular, the  strong pairing interaction for $\eta=0.5$ provides very large strengths for the transfer probabilities. The differences found in the profiles of $P^{Add}(A)$ are due to the following findings: 
The nucleus $^{74}$Cr is predicted almost closed-shell for $\eta=0$ and 1. This leads to negligible values of $P^{Add}$ if Eq. (8) is used. The values obtained with Eq. (10) are obviously larger. For $\eta=0$, the nucleus $^{82}$Cr is also predicted closed-shell (pairing is very weak) and this explains why the value of $P^{Add}$ for $A=80$ obtained with Eq. (8) is larger than that obtained with Eq. (10). For $\eta=1$, both nuclei 
$^{80}$Cr and $^{82}$Cr are predicted closed-shell, so that the two values of $P^{Add}$ at $A=80$, corresponding to Eqs. (8) and (10), are both very close to zero. The case $\eta=0.5$ is different. Pairing correlations are much stronger and pairing is active for all the isotopes. At $A=80$ and 82 pairing is not quenched. Between the results corresponding to Eq. (8) and those corresponding to Eq. (10) there is a shift, that means that the improved formula globally provides an enhancement of pairing correlations.  

Let us now describe Fig. 9. For $\eta=0$ (a) and 0.5 (b), the same remarks as for panel (a) of Fig. 8 may be done. The case $\eta=1$ (c) is different. Like in Fig. 8 (b), we observe that pairing is still active at the drip line (it is not suppressed). There is however a very important difference with respect to panel Fig. 8 (b). A change of slope is found at $A=78$ when Eq. (10) is used. This change of slope is related to a particularly pronounced enhancement of the probability at $A=80$, where the spatial extended $s$ wave function contributes if Eq. (10) is employed. This explains why this change of slope is not found when Eq. (8) is used. The enhancement of the probability for $\eta=1$ is due to the interplay of the spatial extension of the $s$ wave function and the strong surface localization of the pairing interaction. 

By looking at the two figures, one can conclude that results are very strongly case dependent in the specific values of $P^{Add}$. However, there are two general indications that are independent of the intensity of $P^{Add}$ and depend only on the trends. Let us consider only the full lines in the figures, because they correspond to the improved formula for the transfer probability. First, if pairing correlations are quenched at the drip lines (suppression of pairing for weakly bound systems at the drip lines), it is expected that the probability describing the transfer from the nucleus $\tilde{A}$ to the nucleus 
$\tilde{A}+2$ (where $\tilde{A}+2$ is the drip line nucleus) is smaller than the probability associated to the transfer $\tilde{A}-2 \rightarrow \tilde{A}$ (panels (a) and (c) of Fig. 8 and (a) and (b) of Fig. 9). Second, the opportunity to analyze the surface properties of the pairing interaction obviously exists only if pairing survives at the drip line. In this case, a strongly surface-peaked interaction leads to a change of slope in the trends of the transition probabilities (panel (c) of Fig. 9). This special case ($s$ state contributing at the drip line and pairing not suppressed) may thus provide a chance to better understand 
if the pairing interaction is strongly localized at the surface of nuclei or not, by simply looking at the trend of the transfer probabilities. If the change of slope is not found in measurements, one can exclude strongly surface-peaked pairing interactions. As mentioned above, the problem of the persistence of pairing at the drip lines is tightly connected with this. 

Let us now use the parameters of Table III to calculate the transfer probabilities with the pairing interactions that reproduce at best the $S_{2n}$ trends.

\begin{figure}[htb]
\begin{center}
\includegraphics[width=8cm]{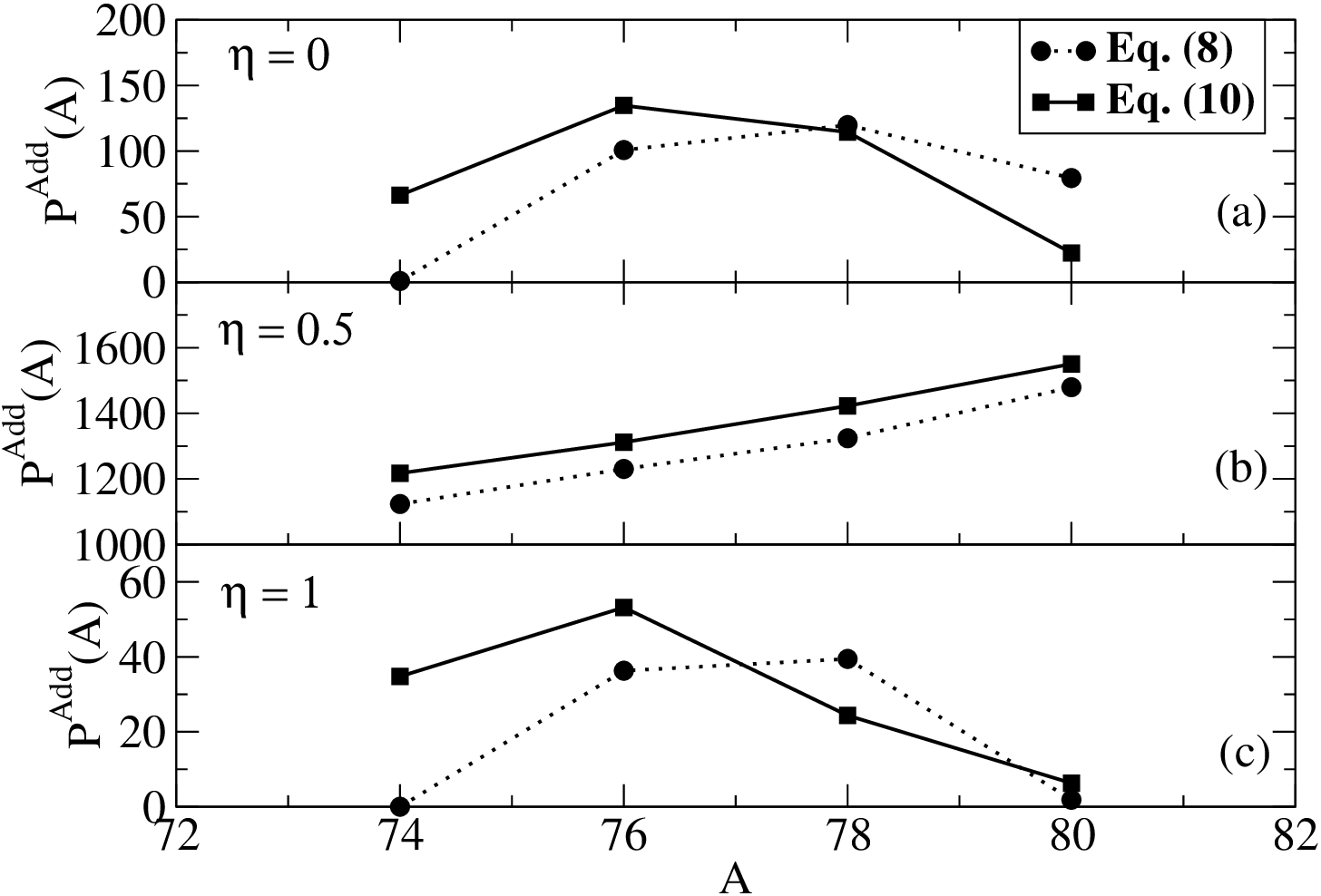}
\end{center}
\caption{Addition transition probabilities for Cr isotopes, from $^{74}$Cr to $^{82}$Cr, for the volume (a), the mixed (b) and the surface-peaked (c) interactions of Table I.}
\end{figure}

\begin{figure}[htb]
\begin{center}
\includegraphics[width=8cm]{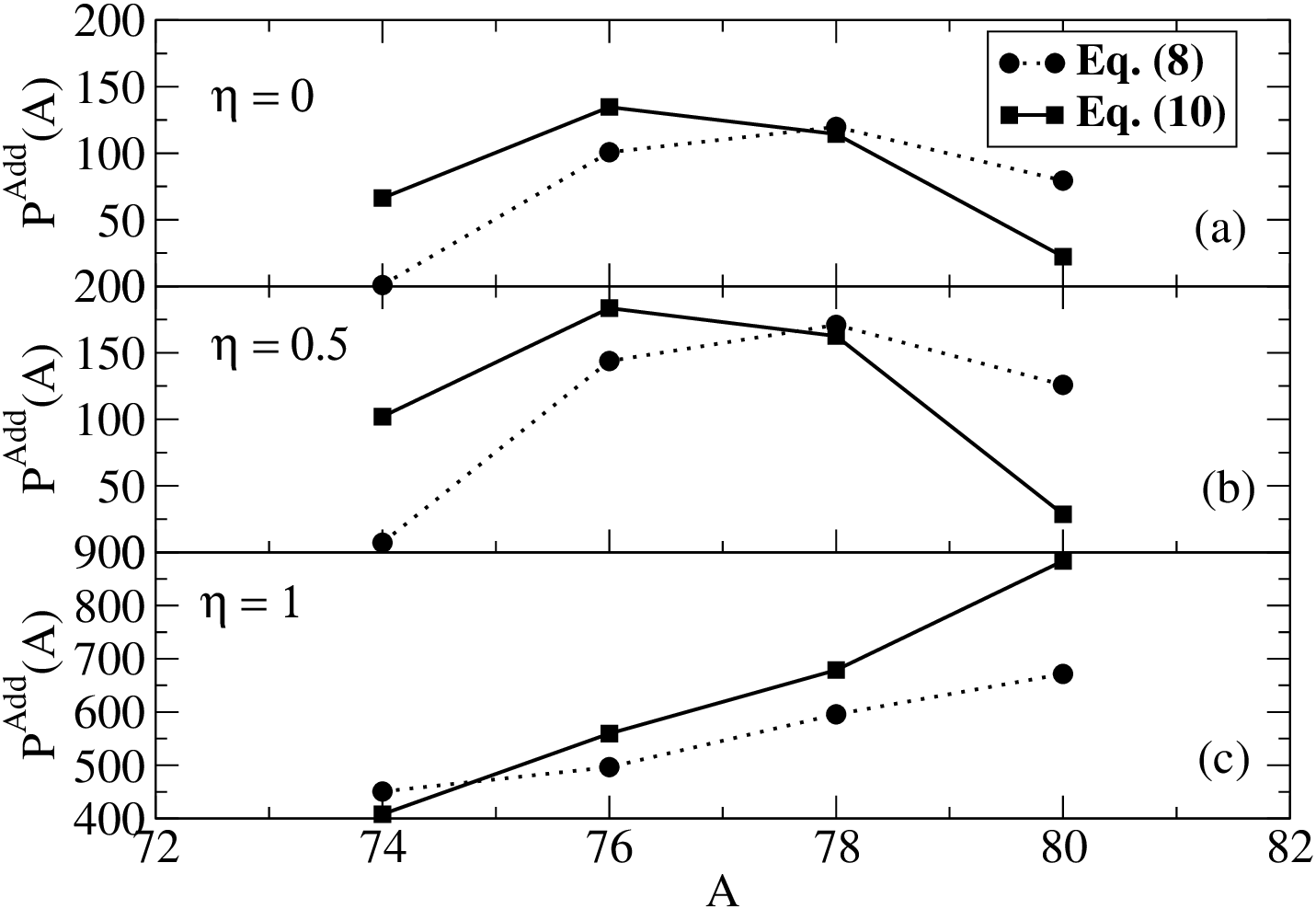}
\end{center}
\caption{Same as in Fig. 8, but with the interactions of Table II.}
\end{figure}

\begin{figure}[htb]
\begin{center}
\includegraphics[width=8cm]{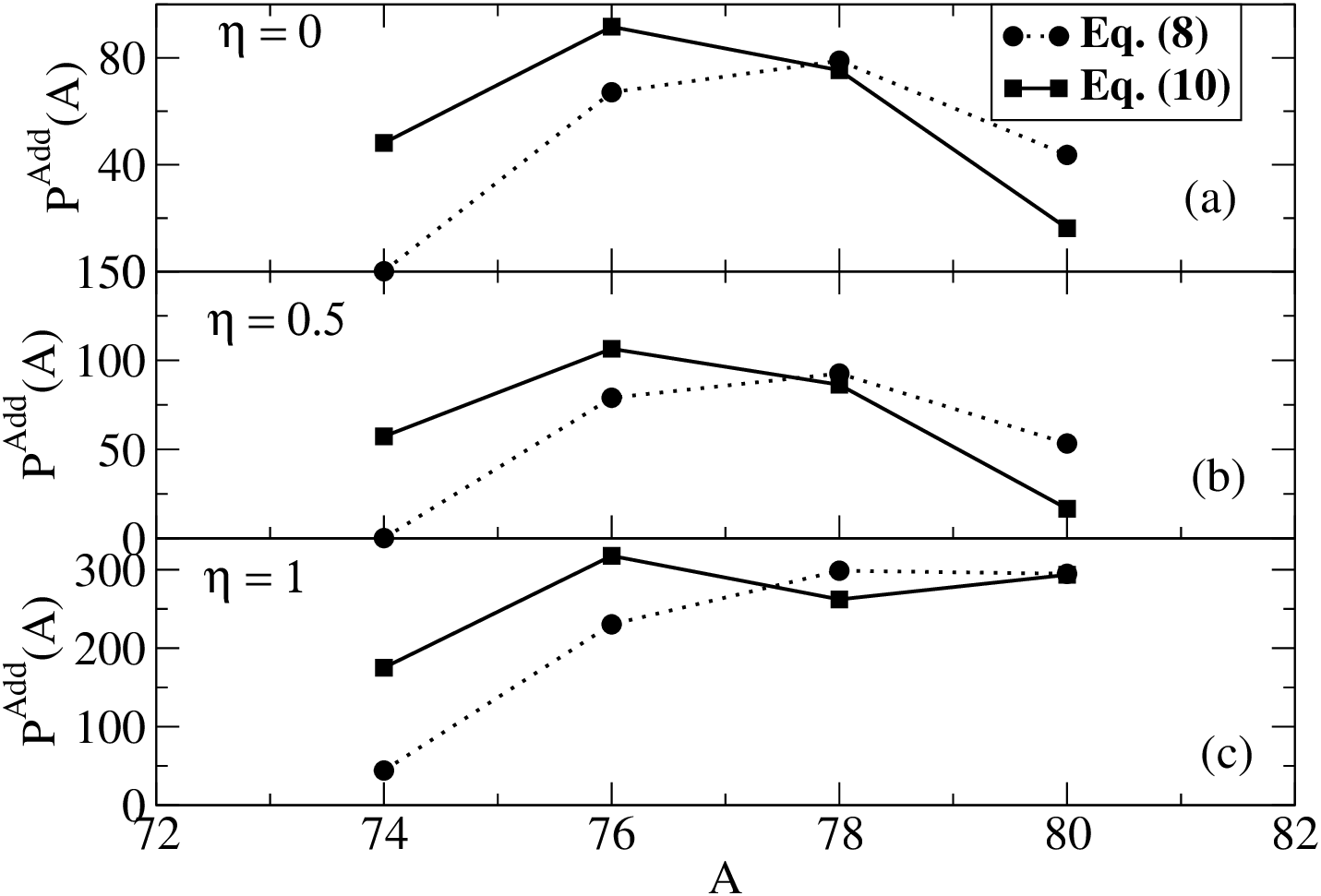}
\end{center}
\caption{Same as in Fig. 8, but with the interactions of Table III.}
\end{figure}

Fig. 10 shows the resulting strengths. The same comments as for Fig. 9 can be done. If pairing persists 
at the drip line, then $P^{Add}(A)$ increases from $A=78$ to $A=80$. Furthermore, a change of slope is predicted in this case for a surface-peaked interaction that amplifies the contribution of the weakly bound $s$ state. 

\section{Conclusions}
We have computed two-neutron transfer probabilities at the (theoretically predicted) neutron drip lines of Cr isotopes, where the last occupied state is the weakly bound neutron $3s1/2$ orbital. We have examined different procedures to adjust a phenomenological pairing interaction in the framework of the HFB model and check the implications of each chosen procedure. Three differently spatially located (volume, mixed and surface-peaked) pairing interactions have been generated for each adjustment procedure. The values of the resulting probabilities are very strongly case dependent. However, a general conclusion can be drawn: if pairing correlations persist at the drip line, a pure surface-peaked interaction would provide not only increasing transfer probabilities from $A=78$ ($^{78}$Cr $\rightarrow$ 
$^{80}$Cr) to $A=80$ ($^{80}$Cr $\rightarrow$ 
$^{82}$Cr), but also a change of slope in the trend at $A=78$. The change of slope originates from an interplay between the spatial extension of the $s$ wave function and the surface localization of the pairing interaction. An interaction that is not surface-peaked leads only to increasing transfer probabilities if pairing is not suppressed. If, on the contrary, pairing is suppressed at the drip line, the transfer probability is expected to decrease going from $A=78$ to $A=80$. 

To conclude, the analysis of Cr isotopes at the drip line can thus offer the opportunity to clarify two aspects: the persistence or suppression of pairing correlations at nuclear drip lines and the surface properties of the pairing interaction. We are of course aware that these isotopes cannot be produced with the present facilities and that, so far, only theoretical studies may be carried out in such region of the nuclear chart, like in Refs. \cite{hagino2} for the odd-even staggering of reaction cross sections.

%
%
%
%-----------------------------------

\end{document}